%% file: main.tex
\documentclass[acmsmall]{acmart}
\usepackage{mathtools,amssymb,nicefrac}
\usepackage{physics}
\usepackage{xspace}
\usepackage{wrapfig}
\usepackage{paralist}
\usepackage{subcaption}
\makeatletter
\newcommand{\pushright}[1]{\ifmeasuring@#1\else\omit\hfill$\displaystyle#1$\fi\ignorespaces}
\makeatother

\allowdisplaybreaks

\let\originalleft\left
\let\originalright\right
\renewcommand{\left}{\mathopen{}\mathclose\bgroup\originalleft}
\renewcommand{\right}{\aftergroup\egroup\originalright}

\input{macros}

\begin{abstract}
In this abstract we study the resource consumption of quantum
programs. Specifically, we focus on the expected runtime of programs
and, inspired by recent methods for probabilistic programs, we develop
a calculus {\em à la} weakest precondition to formally and
systematically derive the (exact) expected runtime of quantum
programs. Notably, the calculus admits a notion of loop runtime
invariant that can be readily used to derive upper bounds of their
runtime. Finally, we show the applicability of our calculus analyzing
the runtime of (a simplified version of) the BB84 quantum key distribution protocol.
\end{abstract}

\begin{document}

\title{Runtime Analysis of Quantum Programs}
\subtitle{A Formal Approach} 
\author{Federico Olmedo}
\affiliation{
 \institution{Department of Computer Science, University of Chile \& IMFD}           
   \city{Santiago}
   \state{RM}
  \country{Chile}   
}
\author{Alejandro D\'{\i}az-Caro}
\affiliation{
  \institution{Universidad Nacional de Quilmes \& ICC (CONICET-UBA)}    
  \streetaddress{Roque S\'aenz Pe\~na 352}
   \city{Bernal}
   \state{Buenos Aires}
   \postcode{B1876BXD}
  \country{Argentina}               
}

\maketitle

\section{Introduction}
For the past decades, quantum programs have been intensively
investigated. Research efforts have ranged from language design and
formal semantics to termination analysis and verification
techniques~\cite{Ying:Book:16}. Nevertheless, the analysis of resource
consumption has received very little attention. 

The most relevant work we are aware of in this direction is the \textsf{Quipper}
system~\cite{Scherer:QIP16:2017}, which compiles quantum programs
described in a high-level language into low-level logical circuits,
and estimates the size of the resulting
circuits, in terms of their number of gates and qbits. The \textsf{ScaffCC}~\cite{Javadi:2014} compilation framework offers similar
possibilities for programs in \textsf{Scaffold}, a high-level
imperative language based on \textsf{C}. %
However, both these systems offer very restricted support for
programs with a recursive control flow. Concretely, they can only encode loops
that are (statically) bounded by a \emph{finite} parameter known at
compilation time, leaving prominent quantum algorithms such as the \textsf{BB84} key distribution
protocol~\cite{Bennett:TCS12:2014} or the Simon's algorithm~\cite{Simon:1997} out
of their scope.

Another related line of research lies in the field of implicit
computational complexity. Dal Lago \etal~\cite{DalLago:TCS:2010} 
have developed a lambda calculus that captures some polynomial time quantum
complexity classes. As such, the lambda calculus can only establish \emph{asymptotic}
guarantees about the behaviour of programs of a particular (polytime) class.

We believe that all players in the quantum programming community can
benefit from more effective tools to estimate the resource consumption of
programs. Indeed, one of the fundamental appeals of quantum computing
is the so-called \emph{quantum speedup}, that is, the possibility that
a quantum computer might efficiently solve problems that are
intractable in the classical world. Having appropriate methods to
formally assess this speedup is thus of utter importance. 

The goal of this work is to provide a first step in this direction. We
specifically focus on the \emph{runtime analysis} of programs and,
inspired by recent methods for probabilistic
programs~\cite{Olmedo:LICS:2016,Kaminski:ESOP:16,Kaminski:JACM:2018},
we develop a calculus {\em à la} weakest precondition to formally and
systematically derive the (exact) runtime of quantum programs
(\S~\ref{sec:ert}). Notably, the calculus admits a notion of loop
\emph{runtime invariant} that can be readily used to derive upper
bounds of their runtime. In comparison to previous works, our calculus
can handle programs with arbitrary loops and is flexible enough to
accomodate different runtime models. We show the applicability of the
calculus analyzing the runtime of an algorithm based on the
\textsf{BB84} quantum key distribution
protocol~\cite{Bennett:TCS12:2014} (\S~\ref{sec:caseStudy}).

We hope that this work serves as starting point for further
developments on the resource consumption of quantum programs; we
briefly discuss some promising directions
in~\S~\ref{sec:discussion}.\footnote{Liu \etal~\cite{Liu:2019}
have independently developed similar ideas to ours to reason about the
expected runtime of quantum programs. While there already exist some differences
between the two approaches, we leave a thorough comparison as future
work since \cite{Liu:2019} became available after the preparation of the current work.}

\section{Programming Model}
Quantum computations are governed by the four postules of the quantum
mechanics:
\begin{description}  
\item[1) State space:] the state of a system is given by a
  density matrix\footnote{A \emph{density matrix} is a square positive
    matrix with trace 1.} acting on a Hilbert space, referred to as
  the system 
  \emph{state space};
\item[2) Evolution:] if in a time lapse a system
  transitions from state $\rho$ to state $\rho'$, then
  $\rho' = \applyM{\rho}{U}$ for some unitary operator $U$;
\item[3) Measurement:] a measurement over a system is modeled by a set
  $\mset = \{M_m\}_{m \in M}$ of operators satisfying the
  normalization condition $\sum_{ m \in M} {M_m}^\dagger M_m = I$.
  The measurement returns an outcome from set
  $M$ and modifies the system state according this outcome: Outcome
  $m \in M$ occurs with probability
  $\condPr{\rho}{\mset}{m} \eqdef \trace({M_{m}}^\dagger
  {M_{m}}\rho)$ and upon this outcome the system transitions
  from state $\rho$ to state
  $\condEff{\rho}{\mset}{m} \eqdef
  (\applyM{\rho}{M_m}) / \condPr{\rho}{\mset}{m}$; 
\item[4) Composition:] the state space of a system composed of several subsystems is the tensor product of the state spaces of its components.
\end{description}

To describe quantum programs we use a core imperative language~\cite{Sanders:MPC:2000}, coined
\qGCL. Programs are defined over a set of
quantum variables (ranged over by $q$). Variable types are interpreted
as Hilbert spaces. Here, we consider only variables of type
$\mathsf{Bool}$ (interpreted as the 2-dimensional Hilbert space
$\CC^2$ with basis $\{ \ket{0}, \ket{1}\}$) and $\mathsf{Int}$
(interpreted as the infinite-dimensional Hilbert space $\CC^\omega$
with basis $\{ \ket{i} \}_{i \in \Int}$). Programs in
\qGCL adhere to the following syntax:
\[
\stmt  \ ::=\  \Skip \ \mid \  \Ass{q}{\ket{b}} \ \mid \  \Ass{\reg{q}}{U \:
       \reg{q}}  \ \mid \   \stmt_1;\stmt_2 \ \mid \  \Choice{\mset[\reg{q}]}{ \overline{m \To \stmt_m}}  \
    \mid \ 
     \WhileDo{(\mset[\reg{q}] = 1)}{\stmt} 
   \]
   \noindent Most language constructs have similar meaning to their
   classical counterpart. $\Skip$ corresponds to a
   no-operation. $\Ass{q}{\ket{b}}$ initializes variable $q$ with
   density operator $\ketbra{b}{b}$, where $\ket b$ lies in the basis
   of its type interpretation.
   $\Ass{\reg{q}}{U \: \reg{q}}$ updates the set of variables $\reg{q}$
   according to the unitary operator $U$. $\stmt_1;\stmt_2$ represents
   the 
   sequential composition of programs $\stmt_1$ and $\stmt_2$.
   $\Choice{\mset[\reg{q}]}{ \overline{m \To \stmt_m}}$ represents the
   quantum counterpart of the traditional $\mathsf{case}$ statement;
   it performs a measurement $\mset$ on variables $\reg{q}$, and
   according to the observed outcome, execution continues with the
   corresponding branch. Finally,
   $\WhileDo{(\mset[\reg{q}] = 1)}{\stmt}$ represents a loop, guarded
   by a binary measurement $\mset$ on $\reg{q}$. Outcome 0 represents
   the loop termination and outcome 1 a further loop iteration.

   The Hilbert state space $\HS_\stmt$ of a program $\stmt$ is given by (the
   class of unit vectors in) the tensor product
   $\bigotimes_{q \in \Var(\stmt)} \HS_q$ of the Hilbert spaces
   associated to each of the variables in $\stmt$. To properly capture
   the semantics of $\stmt$ in case it is non-terminating\footnote{By
     \emph{non-terminating} programs we mean programs that terminate
     with probability less than 1.}, we need to generalize the notion
   of density matrix to that of \emph{partial} density
   matrix\footnote{\emph{Partial} density matrices generalize
     density matrices by allowing traces less or equal than
     1.} \cite{Selinger:MSCS:04}. Then, we interpret $\stmt$ as a transformer
   $\sem{\stmt} \colon \PDO{\HS_\stmt} \To \PDO{\HS_\stmt}$ of partial density
   matrices, where $\PDO{\HS_\stmt}$ denotes the set of all partial density
   matrices over $\stmt$ state space $\HS_\stmt$. Transformer
   $\sem{\stmt}$ is defined by induction on the structure of $\stmt$
   as follows:
   
 $$
\begin{array}{l@{\; \; \; }c@{\; \; \; }l}
  \sem{\Skip}(\rho) &=& \rho\\[0.6ex]
  \sem{\Ass{q}{\ket b}}(\rho)  &=&  \rho\subst{q}{\ket{b}} \\[0.6ex] 
  \sem{\Ass{\reg{q}}{U \: \reg{q}}}(\rho) &=&
                                              \applyM{\rho}{\ext{U}{q}}\\[0.6ex]
  \sem{\stmt_1;\stmt_2}(\rho) &=&\sem{\stmt_1}(\sem{\stmt_2}(\rho))\\[0.6ex]
  \sem{ \Choice{\mset[\reg{q}]}{\overline{m \To \stmt_m}}}(\rho)
                 &=& \sum_m \condPr{\rho}{\mset}{m} \cdot
                   \sem{\stmt_m} (\condEff{\rho}{\mset}{m} )\\[0.9ex]
  \sem{\WhileDo{(\mset[\reg{q}] = 1)}{\stmt}}(\rho)&=& \lfp \bigl(\charFunSem{\mset}{\stmt}\bigr) \\[1.2ex]
\end{array}
$$

\noindent In the second rule, the resulting state
$\rho\subst{q}{\ket{b}}$ is defined as
$\sum_{i \in \{0,1\}} \: \ext{(\ketbra{b}{i})}q \rho \ext{(\ketbra{i}{b})}q$ if
$\type{q}=\mathsf{Bool}$ and as
$\sum_{ i \in \Int } \: \ext{(\ketbra{b}{i})}q\rho\ext{(\ketbra{i}{b})}q$ if
$\type{q}=\mathsf{Int}$, where $\ext Aq$ is the canonical extension of the
matrix $A$ acting on $q$ to the
  dimension of $\rho$. In the third rule, $U$ denotes an unitary operator over
  $\bigotimes_{q \in \reg{q}}\HS_q$. Finally, in the last rule
  $\charFunSem{\mset}{\stmt}(X) \,\eqdef\, \lambda \rho'.\ \condPr{\rho'}{\mset}{0} \cdot
\condEff{\rho'}{\mset}{0} + X\bigl(\condPr{\rho'}{\mset}{1} \cdot
\sem{\stmt}(\condEff{\rho'}{\mset}{1})\bigr)$.

\section{Program Runtimes}\label{sec:ert}

\subsubsection*{Runtime model} Observe that the presence of
measurements endows programs with a probabilistic behaviour: 
programs admit multiple executions, each occurring with a given
probability. Here, we focus on the \emph{expected} or average runtime
of programs, which refers to the weighted sum of the runtime of their
individual executions, where each execution is weighted according to
its probability. In turn, to model the runtime of an individual program execution we
asume that a $\Skip$ statement consumes 1 unit of time and parametrize
the runtime specification of the quantum operations by means of
a function $\timeOf{\cdot}$: a variable initialization with vector
$\ket{b}$ takes $\timeOf{\ket{b}}$ units of time, a state update
induced by a unitary operator $U$ takes $\timeOf{U}$ units of time and
a measurement $\mset$ (together with the modification it induces on
the program state) takes $\timeOf{\mset}$ units of time.

\subsubsection*{Runtime transformer $\ertSymbol$} To formally capture
the expected runtime of programs we use a continuation passing
style, materialized by transformer $\ertSymbol$. If $\stmt$ is a program with
 state space $\HS_\stmt$ and we let $\RT = \PDO{\HS_\stmt} \To \PRealInf$,
 then 
$\ertt{\stmt} \colon \RT \To \RT~,$ and acts as follows: Assume that
$t \colon \RT$ represents the runtime of the program following
$\stmt$, \ie its continuation. Then $\ert{\stmt}{t} \colon \RT$
represents the runtime of $c$ plus its continuation. Here, both
$\ert{\stmt}{t}$ and $t$ have type $\RT$ (rather than simply
$\PRealInf$) because the runtime of programs (in particular, of
$\stmt$ and its continuation) depends on the particular partial
density matrix in $\PDO{\HS_\stmt}$ from which their execution is
started. Finally, observe that to recover the runtime of a plain
program $\stmt$, it suffices to set the runtime of its continuation to
0. Symbolically,
$\ert{\stmt}{\lambda \rho'.\, 0}(\rho)$ gives the runtime of $c$
when executed from initial partial density matrix $\rho$.

The fundamental appeal of this continuation-based approach to model
the runtime of programs is that transforer $\ertSymbol$ admits a
simple and elegant definition by induction on the program structure:
\[
\begin{array}{lcl}
  \ert{\Skip}{t} &=& \lambda \rho.\ 1+  t(\rho) \\[0.7ex]
  \ert{\Ass{q}{\ket b}}{t}  &=&\lambda\rho.\ \timeOf{\ket b} +
                              t\bigl(\rho\subst{q}{\ket{b}}\bigr)\\[0.7ex]  
  \ert{\Ass{\reg{q}}{U \: \reg{q}}}{t} &=&\lambda\rho.\ \timeOf{U} +
                                         t\left(\applyM{\rho}{\ext{U}{q}} \right)\\[0.7ex]
  \ert{\stmt_1;\stmt_2}{t} &=&\ert{\stmt_1}{\ert{\stmt_2}{t}}\\[0.7ex]
  \ert{ \Choice{\mset[\reg{q}]}{\overline{ m \To \stmt_m}}}{t}
                 &=&\lambda\rho.\ \timeOf{\mset} +\sum_m
                   \condPr{\rho}{\mset}{m} \cdot
                   \ert{\stmt_m}{t}
                   \left(\condEff{\rho}{\mset}{m} \right)\\[0.9ex]
  \ert{\WhileDo{(\mset[\reg{q}] = 1)}{\stmt}}{t}&=& \lfp\bigl(\charFun{\mset[\reg{q}]}{\stmt}{t}\bigr)
\end{array}
\]
Here, $\charFun{\mset[\reg{q}]}{\stmt}{t}(t') \,\eqdef\, \lambda\rho. \ \timeOf{\mset}
  + \condPr{\rho}{\mset}{1} \cdot \ert{\stmt}{t'}  (\condEff{\rho}{\mset}{1})+ \condPr{\rho}{\mset}{0}
  \cdot t (\condEff{\rho}{\mset}{0})$ and its least fixed point
  $\lfp\bigl(\charFun{\mset[\reg{q}]}{\stmt}{t}\bigr)$ is taken \wrt the
  pointwise order over $\RT$, \ie $t_1 \preceq t_2$ if $t_1(\rho) \leq t_2(\rho)$ for
  every $\rho$. 

  \subsubsection*{Invariant-based reasoning} Reasoning about the
  runtime of loop-free programs is rather straightforward following
  the rules above. On the contrary, reasoning about the runtime of
  loopy programs requieres determining the least fixed point of
  transformers, which is not a simple task. Nevertheless, if we are
  interested in establishing upper bounds ---rather than exact values---
  for the runtime of loopy programs, we can employ an
  invariant-based argument.
  Concretely,
  \begin{equation}
    \label{eq:inv-rule}
    \charFun{\mset}{\stmt}{t} (I) \preceq I \quad \implies \quad \ert{\WhileDo{(\mset =
      1)}{\stmt}}{t} \preceq I~.
  \end{equation}
 The results follows from a direct
  application of Park's Theorem~\cite{Wechler:MTCS:92}, exploiting $\ertSymbol$ continuity.

\section{Case study}\label{sec:caseStudy}
We demonstrate the applicability of our approach by formally analyzing
the runtime of a simplified version of the BB84 quantum key
distribution algorithm~\cite{Bennett:TCS12:2014}. BB84 is a protocol
to securely create and distribute a shared (\ie symmetric) key between
two parties, say Alice and Bob. Assume the key consists of $m$
bits. To begin with, Alice sends $m$ encoded bits to Bob. To determine
each of these bits, Alice flips two quantum coins; the first coin
determines whether the encoded bit will be $0$ or $1$; the second coin
determines whether she will encode it using basis $\{\ket 0,\ket 1\}$
or $\{\ket +,\ket -\}$. Then Bob continues by measuring each of the
received (encoded) bits. For each of them, he flips a coin to determine the
basis (either $\{\ket 0,\ket 1\}$ or $\{\ket +,\ket -\}$) he will use
for the measurement. Finally, Alice and Bob publish the basis they
employed to respectively encode and measure each bit. Bits whose
respective basis coincide are kept as part of the resulting key; the
remaining bits are discarded. The process continues until completing
the $m$ bits.

\begin{figure}[t!]
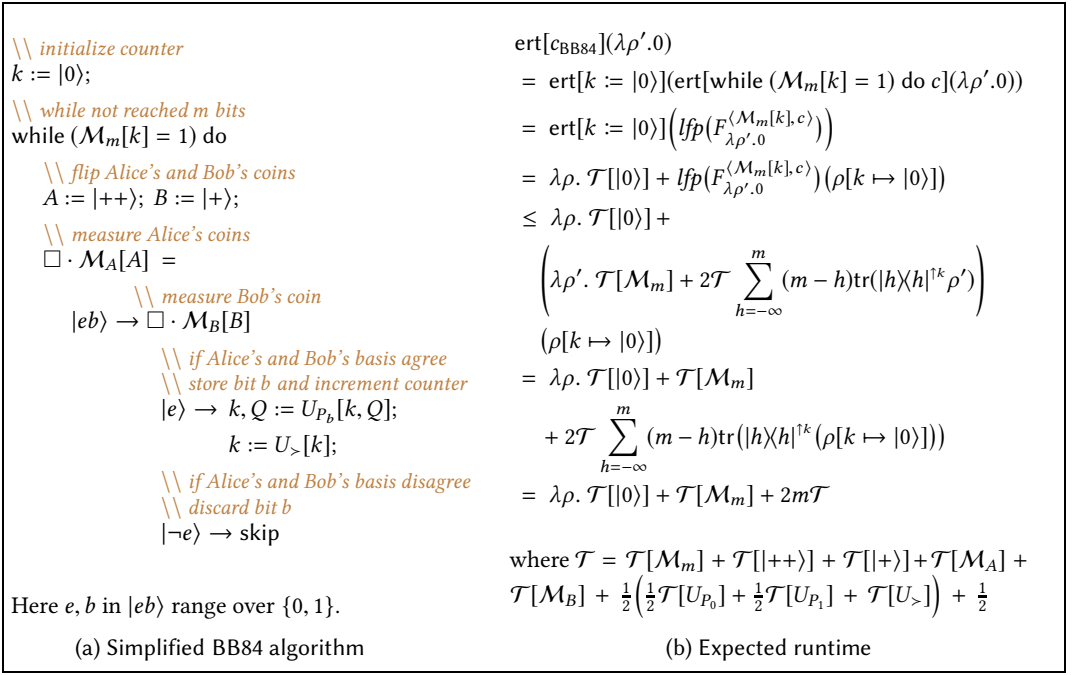

  \fbox{\begin{minipage}{\textwidth}
\begin{subfigure}[b]{.4\textwidth}
   \scalebox{.9}{\begin{minipage}{\textwidth}
  \begin{align*}
   & \codeComment{initialize counter}\\[-1ex]
    & k:=\ket{0};\\
    & \codeComment{while not reached $m$ bits }\\[-1ex]
  &\m{while}\ (\mset_m[k]=1)\ \m{do}\\
  & \begin{aligned}
    \quad & \begin{aligned}
      & \codeComment{flip Alice's and Bob's coins}\\[-1ex]
      & A :=\ket{++}; \ B :=\ket +;\\
    \end{aligned}\\
    & \begin{aligned}
       & \codeComment{measure Alice's coins}\\[-1ex]
      & \square\cdot \mset_A[A] \ = \\
      &\quad
      \begin{aligned}
        \hphantom{\ket{eb}} &\hphantom{\rightarrow } \codeComment{measure Bob's coin}\\[-1ex]
	\ket{eb}&\rightarrow\square\cdot \mset_B[B]\\
	&\qquad
	\begin{aligned}[t]
          &  \codeComment{if Alice's and Bob's basis agree}\\[-1ex]
          &  \codeComment{store bit $b$ and increment counter}\\[-1ex]
	  &\ket e\rightarrow
	  \begin{aligned}[t]
	    &k,Q := U_{P_b}[k,Q];\\
	    &k:=U_{\succ}[k];
	  \end{aligned}\\
          &  \codeComment{if Alice's and Bob's basis disagree}\\[-1ex]
          &  \codeComment{discard bit $b$}\\[-1ex]
	  &\ket{\neg e}\rightarrow\m{skip}
	\end{aligned}\\
      \end{aligned}
    \end{aligned}
  \end{aligned}\\
 \end{align*}
 Here  $e,b$ in $\ket{eb}$ range over $\{0,1\}$.
 \end{minipage}}
 \caption{Simplified BB84 algorithm}
   \label{fig:program}
 \end{subfigure}%
 \hfill
 \begin{subfigure}[b]{.55\textwidth}
   \centering
   \scalebox{.9}{\begin{minipage}{\textwidth}
 \begin{align*}
   & \ert{\stmt_\mathsf{BB84}}{\lambda \rho'. 0}\\
   &=\  \ert{\Ass{k}{\ket 0}}{{\ert{\m{while}\ (\mset_m[k]=1)\ \m{do}\
     \stmt }{\lambda \rho'. 0}}}\\
  &=\  \ert{\Ass{k}{\ket 0}}{\lfp{\bigl( \charFun{\mset_m[k]}{\stmt}{\lambda \rho'. 0} \bigr)}}\\
  &=\ \lambda \rho.\ \timeOf{\ket 0}+ {\lfp{\bigl(
    \charFun{\mset_m[k]}{\stmt}{\lambda \rho'. 0} \bigr)}}\bigl(\rho\subst{k}{\ket{0}}\bigr) \\
  &\leq\  \lambda \rho.\ \timeOf{\ket 0}
\, +\\
   &\quad\left( \lambda \rho'.\ \timeOf{\mset_m}
    + 2 \mathcal T
    \sum_{h=-\infty}^{m}(m-h) 
    {\m{tr}(\ext{\ketbra h}k\rho')}
  \right)\\
&\quad
  \bigl(\rho\subst{k}{\ket{0}}\bigr) \\
  &=\ \lambda \rho.\ \timeOf{\ket 0}+
  \timeOf{\mset_m}\\
&
  \quad+
  2\mathcal T
  \sum_{h=-\infty}^{m}(m-h)  
  {\m{tr}\left(\ext{\ketbra h}k\bigl( \rho\subst{k}{\ket{0}}\bigr)\right)}
  \\
  &=\ \lambda \rho.\ \timeOf{\ket 0}+ \timeOf{\mset_m} + 2m\mathcal T\\[-1ex]
 \end{align*}

 where \(
   \mathcal T  = \:
  \timeOf{\mset_m} \,+\,
  \timeOf{\ket{++}} \,+\, 
  \timeOf{\ket +}
  + 
  \timeOf{\mset_A} \,+\,
  \timeOf{\mset_B} \,+\,
  \frac 12\Bigl(
    \frac 12\timeOf{U_{P_0}}
    +\frac 12\timeOf{U_{P_1}}
  \,+\,\timeOf{U_{\succ}}\Bigr)
  \,+\,\frac 12
\)
\end{minipage}}
 \caption{Expected runtime}
   \label{fig:ertBB84}
 \end{subfigure}
 \end{minipage}}
\caption{Program $\stmt_\mathsf{BB84}$ encoding (a simplified version
  of) the BB84 key distribution algorithm together with the derivation
of its expected runtime.}
 \end{figure}
 
Here we abstract the basis exchange step. We assume that for each bit,
Alice and Bob flip their coins, and immediately
determine whether the basis used by Alice and Bob coincide, keeping or
discarding the bit at hand accordingly. The \qGCL program
$\stmt_{\mathsf{BB84}}$ representing this algorithm is depicted on
Figure~\ref{fig:program}. Variable $k$ (over space $\CC^\omega)$ keeps
track of the number of completed key bits; variable $A$ and $B$ (over
space $\CC^4$ and $\CC^2$, respectively) represent the coin flips of
Alice and Bob, respectively. Finally, variable $Q$ stores (the
successive bits of) the key. Set $\mset_m = \{M_\geq,M_<\}$ measures
counter $k$, distinguishing whether it reached $m$ or not
($M_\geq=\sum_{i=m}^{\infty}\ketbra{i}{i}$ and $M_<=I-M_\geq$). Set
$\mset_A
=\{\ketbra{00}{00},\ketbra{01}{01},\ketbra{10}{10},\ketbra{11}{11}\}$
measures Alice's coin outcomes yielding the encoded bit and the
employed basis.  Finally, set
$\mset_B = \{\ketbra{0}{0},\ketbra{1}{1}\}$ measures Bob's coin
outcome, yielding his employed basis. Unitary operator $U_{P_b}$ acts
on variables $k,Q$ setting the $k$-th bit of $Q$ to $\ket b$ (and
leaving variable $k$ untouched). Finally, unitary operator $U_{\succ}$
acts on variable $k$ by updating its value from $\ket{i}$ to
$\ket{{i+1}}$.

The runtime analysis of the program is depicted in
Figure~\ref{fig:ertBB84}. The crux of the analysis is proving that
runtime
$\lambda \rho'.\ \timeOf{\mset_m} + 2\mathcal T
\sum_{h=-\infty}^{m}(m-h) \cdot {\m{tr}(\ext{\ketbra h}k\rho')}$ is a
fixed point of transformer
$\charFun{\mset_m[k]}{\stmt}{\lambda \rho'. 0}$, where $\stmt$ stands
for the body of $\stmt_{\mathsf{BB84}}$ loop and $\mathcal{T}$ is as
specified in Figure~\ref{fig:ertBB84}; full derivation can be found in the Appendix. The
overall analysis yields that program $\stmt_{\mathsf{BB84}}$ has
(at most) runtime $$\timeOf{\ket 0}+ \timeOf{\mset_m} + 2m\mathcal T$$ for any
initial partial density matrix (from which it is executed), which
matches the anticipated result: in average, the program terminates
after $2m$ iterations; each iteration takes $\mathcal{T}$ units of
time, there is a final measurement upon loop exit requiring
$\timeOf{\mset_m}$ units of time, and the upfront initialization
requires $\timeOf{\ket 0}$ units of time.

\section{Future Work}\label{sec:discussion}
The current work opens several research directions we plan to address
shortly. Among them are \begin{inparaenum}[i)]
  \item defining an operational notion of expected runtime (\eg based
    on quantum Markov chains) and proving
    a correspondence with our approach;
  \item relating the $\ertSymbol$ transformer with the termination behavior of
programs, showing that a finite expected runtime implies termination
with probability 1; and
\item studying the effect of entangled states in the runtime 
  of programs. %
\end{inparaenum}
Another long-term, more challenging goals
comprise \begin{inparaenum}[i)]
  \item studying the problem of automation, more specifically loop
    invariant synthesis;
  \item reasoning about the asymptotic runtime of programs; and
  \item extending the language with further constructs such as general
    recursion or non-determinism.
  \end{inparaenum}

\bibliography{biblio.bib}

\newpage
\appendix
 \section*{Appendix: Omitted calculations from case study in \S~4}
 In this appendix we include the omitted calculations in the runtime
 derivation sketched in Fig.~1b, that is, the application of the proof
 rule in Equation~\eqref{eq:inv-rule} to upper bound the runtime of
 the program loop. We apply the proof rule with runtime invariant
 $$I \ \eqdef \   \lambda \rho.\ \timeOf{\mset_m} + 2\mathcal ( \timeOf{\mset_m} +  \mathcal T^{\star})
 \sum_{h=-\infty}^{m}(m-h) \cdot {\m{tr}(\ext{\ketbra h}k\rho)}~,$$
 where 
 $
   \mathcal T^{\star} \eqdef 
  \timeOf{\ket{++}}+ 
  \timeOf{\ket +}
  + 
  \timeOf{\mset_A} +
  \timeOf{\mset_B}+
  \frac 12\left(
    \frac 12\timeOf{U_{P_0}}
    +\frac 12\timeOf{U_{P_1}}
  +\timeOf{U_{\succ}}\right)
  +\frac 12
$.
The proof rule application requires showing that $I$ is a pre-fixpoint
of runtime transformer
\[
  \charFun{\mset_m[k]}{c}{\lambda \rho'. 0}(t') \ = \ \lambda \rho.\:
  \timeOf{M_m}+\condPr{\rho}{\mset_m}{1} \cdot \ert
    c{t'}\left({\condEff{\rho}{\mset_m}{1}}\right)~.
\]         
(Recall that $\mset_m[k]$ stands for the loop guard measurement and
$\stmt$ for the loop body.) We thus continue unfolding the above
definition calculating $\ert{\stmt}{t'}$. To this end we let 
$ \stmt_1 = A:=\ket{++};B:=\ket +$ be the first two initializations in
$\stmt$, $\stmt_2 = \square\cdot \mset_A[A] \ = \dots$ be the
remaining case statement in $\stmt$ and finally for each
$e,b \in \{0,1\}$, we let
$\stmt_{2eb} = \ket{eb} \to \square\cdot\mset_B[B]\dots$ be the
corresponding branch of $\stmt_2$ case statement. Then we have,

\begin{equation}\label{eq:ertS}
  {\ert\stmt {t'}}
  \ = \ \ert{\stmt_1}{{\ert{\stmt_2}{t'}}}
  \ = \ \ert{\stmt_1}
  {\lambda \rho.\: \timeOf{M_A}+\frac 14
    \sum_{e,b\in\{0,1\}}
    \underline{\ert{\stmt_{2eb}}{t'}}\left(\condEff{\rho}{\mset_A}{eb}\right)}
\end{equation}
where
\begin{align*}
  &\underline{\ert{\stmt_{2eb}}{t'}}\\
  & =
    \lambda \rho.\: \timeOf{M_B}
    +\frac 12.\ert{[k,Q]:=U_{P_b}[k,Q]; k:=U_{\succ}\ k}{t'}
    \left(\condEff{\rho}{\mset_B}{e}\right)
    +\frac 12.\underline{\ert{\m{skip}}{t'}\left(\condEff{\rho}{\mset_B}{\neg e}\right)}\\
  & =
    \lambda \rho. \:\timeOf{M_B}
    +\frac 12.\underline{\ert{[k,Q]:=U_{P_b}[k,Q]; k:=U_{\succ}\ k}{t'}}
    \left(\condEff{\rho}{\mset_B}{e}\right)
    +\frac 12.\left(1+{t'}\left(\condEff{\rho}{\mset_B}{\neg e}\right)\right)
  \\
  & =
    \lambda \rho. \:\timeOf{M_B}
    +\frac 12.\ert{[k,Q]:=U_{P_b}[k,Q]}{\underline{\ert{k:=U_{\succ}\ k}{t'}}}
    \left(\condEff{\rho}{\mset_B}{e}\right)
    +\frac 12.\left(1+{t'}\left(\condEff{\rho}{\mset_B}{\neg e}\right)\right)
  \\
  & =
    \lambda \rho. \:\timeOf{M_B}
    +\frac 12.\underline{\ert{[k,Q]:=U_{P_b}[k,Q]}
    {\lambda \rho. \:\timeOf{U_{\succ}}+t' (\ext{U_{\succ}}k\rho{\ext{U_{\succ}}k}^\dagger)}}
    \left(\condEff{\rho}{\mset_B}{e}\right)
    \\
  &
    \pushright{
    +\frac 12.\left(1+{t'}\left(\condEff{\rho}{\mset_B}{\neg e}\right)\right)
    }
  \\
  & = 
    \lambda \rho. \:\timeOf{M_B}
    +\frac 12.
    \left(
    \lambda \rho'. \:\timeOf{U_{P_b}}+
    \underline{\left( {\lambda \rho.\timeOf{U_{\succ}}+t' (\ext{U_{\succ}}k\rho{\ext{U_{\succ}}k}^\dagger)}\right) \left( \ext{U_{P_b}}{k,Q}\rho'{\ext{U_{P_b}}{k,Q}}^\dagger\right)}
    \right)
    \left(\condEff{\rho}{\mset_B}{e}\right)
  \\
  &\pushright{
    +\frac 12.\left(1+{t'}\left(\condEff{\rho}{\mset_B}{\neg e}\right)\right)
    }
  \\
  & = 
    \lambda \rho. \:\timeOf{M_B}
    +\frac 12. 
    \underline{\left(
    \lambda \rho'. \:\timeOf{U_{P_b}}+
    \timeOf{U_{\succ}}+t' (\ext{U_{\succ}}k
    \ext{U_{P_b}}{k,Q}\rho'{\ext{U_{P_b}}{k,Q}}^\dagger
    {\ext{U_{\succ}}k}^\dagger) 
    \right)
    \left(\condEff{\rho}{\mset_B}{e}\right)}
  \\
  &\pushright{
    +\frac 12.\left(1+{t'}\left(\condEff{\rho}{\mset_B}{\neg e}\right)\right)
    }
  \\
  & = 
    \lambda \rho. \:\timeOf{M_B}
    +\frac 12. 
    \left(
    \timeOf{U_{P_b}}+
    \timeOf{U_{\succ}}+t' (\ext{U_{\succ}}k
    \ext{U_{P_b}}{k,Q}\left(\condEff{\rho}{\mset_B}{e}\right){\ext{U_{P_b}}{k,Q}}^\dagger
    {\ext{U_{\succ}}k}^\dagger) 
    \right)
    +\frac 12.\left(1+{t'}\left(\condEff{\rho}{\mset_B}{\neg e}\right)\right)
\end{align*}

Plugging the above expansion of $\ertt{\stmt_{2eb}}$ into
\eqref{eq:ertS} yields:
 \begin{align*}
   &\ert{\stmt_1}
     {\lambda \rho. \:\timeOf{M_A}+\frac 14
     \sum_{e,b\in\{0,1\}}
     \underline{\ert{\stmt_{2eb}}{t'}\left(\condEff{\rho}{\mset_A}{eb}\right)}}
  \\
  &=
    \ert{\stmt_1}{
    \begin{aligned}
      \lambda \rho. \:\timeOf{M_A}+
    &\underline{\frac 14\sum_{e,b\in\{0,1\}}
    \left( 
    \timeOf{M_B}\right.}\\
     &\underline{+\frac 12. 
      \left(
        \timeOf{U_{P_b}}+
        \timeOf{U_{\succ}}+t' ({\ext{U_{\succ}}k
          \ext{U_{P_b}}{k,Q}
          \left( \condEff{\left( \condEff{\rho}{\mset_A}{eb}\right)}{\mset_B}{e}\right)
          {\ext{U_{P_b}}{k,Q}}^\dagger
          {\ext{U_{\succ}}k}^\dagger}) 
      \right)}\\
    &
      \left.
    \underline{
        +\frac 12.
        \left(
          1+{t'}
          \left(
            \condEff{
              \left(
                \condEff{\rho}{\mset_A}{eb}
              \right)
            }{\mset_B}{\neg e}
          \right)
        \right)
      }
    \right)
    \end{aligned}
        }
  \\
  &=
    \ert{\underline{\stmt_1}}{
    \begin{aligned}
    &\lambda \rho. \: \timeOf{M_A}
    +
    \timeOf{M_B}+\frac 12.(
    \frac 12.\timeOf{U_{P_0}}
    +\frac 12.\timeOf{U_{P_1}}
    +\timeOf{U_{\succ}})
    +\frac 12
    \\
 &
    +\frac 18
     \sum_{e,b\in\{0,1\}}
     \left(
      t' \left({\ext{U_{\succ}}k
        \ext{U_{U_{P_b}}}{k,Q}
        \left( \condEff{\left( \condEff{\rho}{\mset_A}{eb} \right)}{\mset_B}{e}\right)
        {\ext{U_{U_{P_b}}}{k,Q}}^\dagger
        {\ext{U_{\succ}}k}^\dagger}\right) 
    +{t'}
    \left( \condEff{\left( \condEff{\rho}{\mset_A}{eb} \right)}{\mset_B}{\neg e}\right)
  \right)
\end{aligned}
   }
  \\
  &=
    \ert{\underline{A:=\ket{++}}}{
    \begin{aligned}
      &\lambda \rho. \: \timeOf{\ket +}
      +
        \timeOf{M_A} +
        \timeOf{M_B}+\frac 12.(
        \frac 12.\timeOf{U_{P_0}}
        +\frac 12.\timeOf{U_{P_1}}
        +\timeOf{U_{\succ}})
        +\frac 12\\
      &
      +\frac 18
      \sum_{e,b\in\{0,1\}}
      \begin{aligned}[t]
        &
       \left( 
        t' \left({\ext{U_{\succ}}k
          \ext{U_{U_{P_b}}}{k,Q}
        \left( \condEff{\left( \condEff{\left(\rho[B\mapsto \ketbra{+}{+}]\right)}{\mset_A}{eb} \right)}{\mset_B}{e}\right)
          {\ext{U_{U_{P_b}}}{k,Q}}^\dagger
          {\ext{U_{\succ}}k}^\dagger}\right) \right.
        \\
        &\left.+{t'}
    \left( \condEff{\left( \condEff{\left(\rho[B\mapsto \ketbra{+}{+}]\right)}{\mset_A}{eb} \right)}{\mset_B}{\neg e}\right)
          \right)
      \end{aligned}
    \end{aligned}
          }
  \\
  &=
    \lambda \rho.\:  \mathcal T^{\star}+\frac 18 \sum_{e,b\in\{0,1\}} \left(t' {\psi_{eb}(\rho)}+t' {\phi_{eb}(\rho)}\right)
 \end{align*}
 where
 \begin{itemize}
 \item $\mathcal T^{\star}$ is as already defined on the previous page,
 \item $\psi_{eb}(\rho) = {{\ext{U_{\succ}}k \ext{U_{U_{P_b}}}{k,Q} \left(
         \condEff{\left( \condEff{\left(\rho[AB\mapsto
                 \ketbra{+++}{+++}]\right)}{\mset_A}{eb}
           \right)}{\mset_B}{e}\right) {\ext{U_{U_{P_b}}}{k,Q}}^\dagger
       {\ext{U_{\succ}}k}^\dagger}}$, and 
 \item $\phi_{eb}(\rho) = {\condEff{\left( \condEff{\left(\rho[AB\mapsto \ketbra{+++}{+++}]\right)}{\mset_A}{eb} \right)}{\mset_B}{\neg e}}$.
 \end{itemize}
 
 \bigskip
 
 \noindent All in all, we have:
\begin{align*}
  \charFun{\mset_m[k]}{\stmt}{\lambda \rho'. 0}(t')
  \ = \ \lambda \rho.\: \timeOf{M_m}
  &+\condPr{\rho}{\mset_m}{1}. \left\{ \mathcal T^{\star} + \frac 18 \sum_{e,b\in\{0,1\}} \left(t' {\psi_{eb}\left(\condEff{\rho}{\mset_m}{1}\right)} +{t'} \phi_{eb}\left(\condEff{\rho}{\mset_m}{1}\right) \right) \right\}\\
\end{align*}

Now we proceed to verify that $I$ is a (pre-)fixpoint of
$\charFun{\mset_m[k]}{\stmt}{\lambda \rho'. 0}$:

\begin{align*}
  &\underline{\charFun{\mset_m[k]}{\stmt}{\lambda \rho'. 0}(I)}\\
  &\com{definition of $\charFun{\mset_m[k]}{\stmt}{\lambda \rho'. 0}$}\\
  &=\begin{aligned}[t]
    \lambda \rho.\:
    \timeOf{M_m}
    &+
    \condPr{\rho}{\mset_m}{1}
    \left\{
    \mathcal T^{\star}
    +
    \frac 18
    \sum_{e,b\in\{0,1\}}
    \left(
    \underline{I\psi_{eb}\left(\condEff{\rho}{\mset_m}{1}\right)}
    +
    \underline{I\phi_{eb}\left(\condEff{\rho}{\mset_m}{1}\right)}
    \right)
  \right\}
   \end{aligned}
  \\[1ex]
  &\com{definition of $I$ and application}\\[1ex]
  &=
    \begin{aligned}[t]
     \lambda \rho.\:
     \timeOf{M_m}
     &+
     \condPr{\rho}{\mset_m}{1}
     \left\{
     \mathcal T^{\star}
     +
     \frac 18 
     \sum_{e,b\in\{0,1\}}
    \right.
    \\
     &
     \left(
     \timeOf{M_m}
     +
     2(\timeOf{M_m}+\mathcal T^{\star})
     \sum_{h=-\infty}^{m}(m-h)\m{tr}\left(\ext{\ketbra h}k\psi_{eb}\left(\condEff{\rho}{\mset_m}{1}\right)\right)
     \right.
   \\
   &
     \left.
     \left.
     +
     \timeOf{M_m}
     +
     2(\timeOf{M_m}+\mathcal T^{\star})
     \underline{\sum_{h=-\infty}^{m}(m-h)\m{tr}(\ext{\ketbra h}k\phi_{eb}\left(\condEff{\rho}{\mset_m}{1}\right))}
     \right)
     \right\}
     \end{aligned}
  \\[1ex]
  &\com{$
    \sum_{h=-\infty}^{m}(m-h)\m{tr}(\ext{\ketbra h}k\phi_{eb}\left(\condEff{\rho}{\mset_m}{1}\right))
    =
    \sum_{h=-\infty}^{m-1}(m-h)\frac{\m{tr}(\ext{\ketbra h}k\rho)}{\condPr{\rho}{\mset_m}{1}}
    $}\\[1ex]
  &=
    \begin{aligned}[t]
     \lambda \rho.\:
     \timeOf{M_m}
     &+
     \condPr{\rho}{\mset_m}{1}
     \left\{
     \mathcal T^{\star}
     +
     \frac 18 
     \sum_{e,b\in\{0,1\}}
    \right.
    \\
     &
     \left(
     \timeOf{M_m}
     +
     2(\timeOf{M_m}+\mathcal T^{\star})
     \underline{\sum_{h=-\infty}^{m}(m-h)\m{tr}\left(\ext{\ketbra h}k\psi_{eb}\left(\condEff{\rho}{\mset_m}{1}\right)\right)}
     \right.
   \\
   &
     \left.
     \left.
     +
     \timeOf{M_m}
     +
     2(\timeOf{M_m}+\mathcal T^{\star})
     \sum_{h=-\infty}^{m-1}(m-h)\frac{\m{tr}(\ext{\ketbra h}k\rho)}{\condPr{\rho}{\mset_m}{1}}
     \right)
     \right\}
     \\
     \end{aligned}
  \\[1ex]
  &\com{$
    \sum_{h=-\infty}^{m}(m-h)\m{tr}(\ext{\ketbra h}k\psi_{eb}\left(\condEff{\rho}{\mset_m}{1}\right))
    =
    \sum_{h=-\infty}^{m}(m-h)\frac{\m{tr}(\ext{\ketbra{h-1}}k\rho)} {\condPr{\rho}{\mset_m}{1}}
    $}\\[1ex]
  &=
    \begin{aligned}[t]
     \lambda \rho.\:
     \timeOf{M_m}
     &+
     \condPr{\rho}{\mset_m}{1}
     \left\{
     \mathcal T^{\star}
     +
     \underline{\frac 18 \sum_{e,b\in\{0,1\}}}
    \right.
    \\
     &
     \left(
     \timeOf{M_m}
     +
     2(\timeOf{M_m}+\mathcal T^{\star})
     {\sum_{h=-\infty}^{m}(m-h)\frac{\m{tr}(\ext{\ketbra{h-1}}k\rho)} {\condPr{\rho}{\mset_m}{1}}}
     \right.
   \\
   &
     \left.
     \left.
     +
     \timeOf{M_m}
     +
     2(\timeOf{M_m}+\mathcal T^{\star})
     \sum_{h=-\infty}^{m-1}(m-h)\frac{\m{tr}(\ext{\ketbra h}k\rho)}{\condPr{\rho}{\mset_m}{1}}
     \right)
     \right\}
     \\
     \end{aligned}
  \\[1ex]
  &\com{Removing the summation}\\[1ex]
  &=
    \begin{aligned}[t]
     \lambda \rho.\:
     \timeOf{M_m}
     &+
     \condPr{\rho}{\mset_m}{1}
     \left\{
     \mathcal T^{\star}
     +
     \underline{\frac 12}
    \right.
    \\
     &
     \left(
     \underline{\timeOf{M_m}
     +
     2(\timeOf{M_m}+\mathcal T^{\star})
     {\sum_{h=-\infty}^{m}(m-h)\frac{\m{tr}(\ext{\ketbra{h-1}}k\rho)} {\condPr{\rho}{\mset_m}{1}}}}
     \right.
   \\
   &
     \left.
     \left.
     \underline{+
     \timeOf{M_m}
     +
     2(\timeOf{M_m}+\mathcal T^{\star})
     \sum_{h=-\infty}^{m-1}(m-h)\frac{\m{tr}(\ext{\ketbra h}k\rho)}{\condPr{\rho}{\mset_m}{1}}}
     \right)
     \right\}
     \\
     \end{aligned}
  \\[1ex] 
  &\com{Adding up}\\[1ex]
  &=
    \begin{aligned}[t]
     \lambda \rho.\:
     \timeOf{M_m}
     &+
     \underline{\condPr{\rho}{\mset_m}{1}}
     \left\{
     \underline{\mathcal T^{\star}
     +
     \timeOf{M_m}
     +
     (\timeOf{M_m}+\mathcal T^{\star})}
    \right.
    \\
     &
     \left(
     {\sum_{h=-\infty}^{m}(m-h)\frac{\m{tr}(\ext{\ketbra{h-1}}k\rho)} {\condPr{\rho}{\mset_m}{1}}}
     \left.
     +
     \sum_{h=-\infty}^{m-1}(m-h)\frac{\m{tr}(\ext{\ketbra h}k\rho)}{\condPr{\rho}{\mset_m}{1}}
     \right)
     \right\}
     \\
     \end{aligned}
  \\[1ex]
  &\com{Factoring $(\timeOf{M_m}+\mathcal T^{\star})$ out and distributing
    $\condPr{\rho}{\mset_m}{1}$}\\[1ex]
  &=
    \begin{aligned}[t]
     \lambda \rho.\:
     \timeOf{M_m}
     &+
     (\timeOf{M_m}+\mathcal T^{\star})
    \\
     &
     \left\{
     \underline{\condPr{\rho}{\mset_m}{1}}
     +
     \sum_{h=-\infty}^{m}(m-h){\m{tr}(\ext{\ketbra{h-1}}k\rho)}
     +
     \sum_{h=-\infty}^{m-1}(m-h){\m{tr}(\ext{\ketbra h}k\rho)}
     \right\}
     \\
     \end{aligned}
  \\[1ex]
  &\com{
    $
    \condPr{\rho}{\mset_m}{1}
    =
    \sum_{h=-\infty}^m \m{tr}(\ext{\ketbra h}k\rho)$
    }
  \\[1ex]
  &=
    \begin{aligned}[t]
    \lambda \rho.\:
    \timeOf{M_m}
    &+
    (\timeOf{M_m}+\mathcal T^{\star})
    \\
     &
    \left\{
    \sum_{h=-\infty}^m \m{tr}(\ext{\ketbra h}k\rho)
    +
    \sum_{\underline{h=-\infty}}^{m}(m-\underline h)
    {\m{tr}(\ext{\ketbra{\underline{h-1}}}k\rho)}
    +
    \sum_{h=-\infty}^{m-1}(m-h){\m{tr}(\ext{\ketbra h}k\rho)}
    \right\}
     \\
     \end{aligned}
  \\[1ex]
  &\com{Summing up}
  \\[1ex]
  &=
    \lambda \rho.\:
    \timeOf{M_m}
    +
    2
    (\timeOf{M_m}+\mathcal T^{\star})
    {\sum_{h=-\infty}^{m}(m-h){\m{tr}(\ext{\ketbra h}k\rho)}}
  \\[1ex]
  &\com{Definition of $I$}
  \\[1ex]
  &= I
 \end{align*}
 
\end{document}

%% file: macros.tex
\newcommand{\ie}{i.e.\@\xspace}

\newcommand{\wrt}{w.r.t.\@\xspace}
\newcommand{\eg}{e.g.\@\xspace}

\newcommand{\etal}{et al.\@\xspace}

\newcommand{\PRealInf}{\ensuremath{\mathbb R_{\geq 0}^\infty}\xspace}

\newcommand{\Int}{\ensuremath{\mathbb Z}\xspace}

\newcommand{\Skip}{\mathsf{skip}}

\newcommand{\Ass}[2]{#1 \coloneqq #2}
\newcommand{\Choice}[2]{\square\  #1 \,=\,  #2}
\newcommand{\While}{\mathsf{while}}
\newcommand{\Do}{\mathsf{do}}
\newcommand{\WhileDo}[2]{{\While} \:\allowbreak #1 \: \Do \, \allowbreak #2}

\newcommand{\Var}{\mathit
  {Var}}  
\newcommand{\reg}[1]{\overline{#1}}   

\newcommand{\codeComment}[1]{\textcolor{brown}{\small
    \textbackslash\textbackslash\ \textsl{#1}}}
\newcommand{\stmt}{c}   
\newcommand{\qGCL}{\ensuremath{\mathsf{qGCL}}\xspace}   

\newcommand{\condEff}[3]{\left. #1 \right|_{#2 = #3}}
\newcommand{\condPr}[3]{\Pr{#1}{#2 {=} #3}}
\newcommand{\applyM}[2]{{#2}{#1}{#2}^\dagger}
\newcommand{\sem}[1]{\llbracket #1 \rrbracket}       
\newcommand{\type}[1]{\mathit{type}\left(#1\right)}     

\newcommand\m[1]{\mathsf{#1}}
\newcommand{\charFunSem}[2]{\Phi^{\langle #1,#2 \rangle}}
\newcommand\com[1]{\fbox{\small\em #1}}

\providecommand{\eqdef}{\circeq}
\newcommand{\To}{\rightarrow}                          

\newcommand{\subst}[2]{[#1 \mapsto #2]}
\newcommand{\lfp}{\mathit{lfp}}

\newcommand{\prsymbol}{\mathsf{Pr}}
\renewcommand{\Pr}[2]{\prsymbol_{#1}\left[ #2\right]}

\newcommand{\HS}{\ensuremath{\mathcal{H}}\xspace}   
\newcommand{\CC}{\ensuremath{\mathbb{C}}\xspace}   
\newcommand{\PDO}[1]{\mathcal{P}_{#1}}    

\newcommand{\mset}{\mathcal{M}}   

\newcommand\ext[2]{#1^{\scriptscriptstyle\uparrow #2}}

\newcommand{\ert}[2]{\ertt{#1}\left(#2\right)}
\newcommand{\ertt}[1]{\ertSymbol\left[#1\right]}
\newcommand{\ertSymbol}{\mathsf{ert}}
\newcommand{\timeOf}[1]{\mathcal{T}[#1]}
\newcommand{\charFun}[3]{F^{\langle #1,#2 \rangle}_{#3}}
\newcommand{\RT}{\ensuremath{\mathbb{T}}}
